# An Approach to Sparse Continuous-time System Identification from Unevenly Sampled Data

**Rui Teixeira Ribeiro**[1*]    **Alexandre Mauroy**[1]    **Jorge Goncalves**[1]

[1]*Luxembourg Centre for Systems Biomedicine, University of Luxembourg*

## Abstract

In this work, we address the problem of identifying sparse continuous-time dynamical systems when the spacing between successive samples (the sampling period) is not constant over time.

The proposed approach combines the leave-one-sample-out cross-validation error trick from machine learning with an iterative subset growth method to select the subset of basis functions that governs the dynamics of the system. The least-squares solution using only the selected subset of basis functions is then used. The approach is illustrated on two examples: a 6-node feedback ring and the Van der Pol oscillator.

## 1. Introduction

Inferring the governing equations of dynamical systems from data is a key challenge in many fields of science and engineering. In this work, we address the problem of identifying sparse continuous-time dynamical systems when the spacing between successive samples (the sampling period) is not constant over time.

---

[*] *Corresponding author: rui.t.ribeiro@gmail.com*

To identify continuous-time dynamical systems, two types of approaches are typically used: a) *indirect approach*, where an equivalent discrete-time system is first inferred from samples and is then converted to a continuous-time system, e.g., [1], [2]; b) *direct approach*, where the continuous-time dynamics are directly inferred from samples, e.g., [3], [4]. Since we are considering a scenario where the sampling period changes over time, indirect approaches (where sampling frequency is assumed constant) are not appropriate.

On the other hand, most physical and biochemical systems have only a few relevant terms that define their dynamics, making the governing equations sparse in the space of possible functions (*basis function*) [5]. While inferring such systems, it is therefore important to use sparsity promoting techniques to avoid overfitting.

Several methods have been proposed in the literature to promote sparsity. For example, some methods use regularization techniques, such as LASSO [6] or reweighted $\ell_1$-norm minimization [7], or penalise model complexity using AIC or BIC criterion [8]. Others are based on *iterative subset selection* of the basis functions. For example, in [5], a subset pruning approach is used where, starting by the full set, some terms that govern the dynamics are iteratively removed until a stopping criterion is met.

Belonging to the iterative subset selection methods, the proposed approach combines the leave-one-sample-out cross-validation (LOOCV) error trick from machine learning with an iterative subset growth method to select the subset of basis functions that governs the dynamics of the system. The least-squares solution using only the selected subset of basis functions is then used. In section 3, we illustrate the approach on two examples: a 6-node feedback ring and the Van der Pol oscillator.

## 2. Method

Consider a dynamical system of the form

$$\dot{x} = F(x), \quad x \in \mathbb{R}^n$$

where $x = [x_1 \ x_2 \ \cdots \ x_n]^T$ represents the state variables of the system and $F(x)$ is assumed to be a linear combination of non-linear functions of the state variables, known as *basis functions*. We can write

$$\dot{x} = Z\theta$$

where $\theta = [\theta_1 \ \theta_2 \ \cdots \ \theta_p]^T$ are the basis functions ($p$ denotes the total number of possible basis functions) and the coefficients in matrix $Z \in \mathbb{R}^{n \times p}$ represent the weights of each basis function for each state variable,

$$Z = \begin{bmatrix} | & | & & | \\ \zeta_1 & \zeta_2 & \cdots & \zeta_n \\ | & | & & | \end{bmatrix}^T$$

Note that the system is assumed to be sparse and, thus, matrix $Z$ consists of only a few non-zero elements. The goal is to infer $Z$ from data, given a set of possible basis functions $\theta$. Finding the linear combination of basis functions for each state variable $x_i$ can be posed in terms of a least-squares problem,

$$\dot{x}_i(t) = \Theta(t)\,\zeta_i$$

where each column of $\Theta(t) \in \mathbb{R}^{m \times p}$ represents the values of a basis function at time $t = t_0, t_1, \cdots, t_{m-1}$. The total number of samples is denoted by $m$.

$$\Theta(t) = \begin{bmatrix} | & | & & | \\ \theta_1(t) & \theta_2(t) & \cdots & \theta_p(t) \\ | & | & & | \end{bmatrix}$$

A key point in this approach is that the sampling frequency is not assumed to be constant over time and, therefore, data is sampled at time $t = t_0, t_1, \cdots, t_{m-1}$ not evenly spaced.

As only data $x(t)$ is usually available, the time derivatives of the state variables are estimated by means of cubic spline interpolation [9]†.

## 2.1. System Identification based on Subset Growth (SISuG)

To obtain a sparse vector $\zeta_i$, a least-squares approach imposing sparsity should be employed. Herein, we propose an approach that evolves in the direction of maximal to minimal sparsity while the number of non-zero coefficients in $\zeta_i$ (i.e., the number of considered basis functions) is iteratively increased until a stopping criterion is met. Figure 1 shows the block diagram of the proposed approach to **s**ystem **i**dentification based on **su**bset **g**rowth (SISuG).

For each state variable $x_i$, the number of non-zero coefficients in vector $\zeta_i$ (denoted by $k$) is gradually increased (from $k = 1$ to $k = p$) while a stopping criterion is not satisfied.

For each $k$, all possible combinations of basis functions are examined and the leave-one-sample-out cross-validation (LOOCV) error is used to compute the respective prediction errors (details in section 2.2). The minimum prediction error for each $k$ is denoted by $\varepsilon_k$.

To make the approach computationally faster, the LOOCV error trick was employed (see section 2.2), meaning that the prediction error computation is performed only once for each combination of basis functions, instead of $m$ times ($m$ denotes the number of samples).

The iterative process stops when the minimum prediction error $\varepsilon_k$ does not decrease by more than one order of magnitude below the previous minimum prediction error $\varepsilon_{k-1}$, i.e. the stopping criterion is given by

$$\varepsilon_{k-1} - \varepsilon_k \leq 0.1 \varepsilon_{k-1}$$

The least-squares solution using the combination of basis functions that corresponds to $\varepsilon_{k-1}$ is then used. The solution is denoted by $\widehat{\mathbf{Z}}$.

---

† We used the *spline* algorithm available on *GNU Octave* (version 4.2.1) to fit cubic splines (piecewise cubic polynomials) to the data points. Next, the derivative of each cubic polynomial was straightforwardly computed since the derivative of $ax^3 + bx^2 + cx + d$ equals $3ax^2 + 2bx + c$.

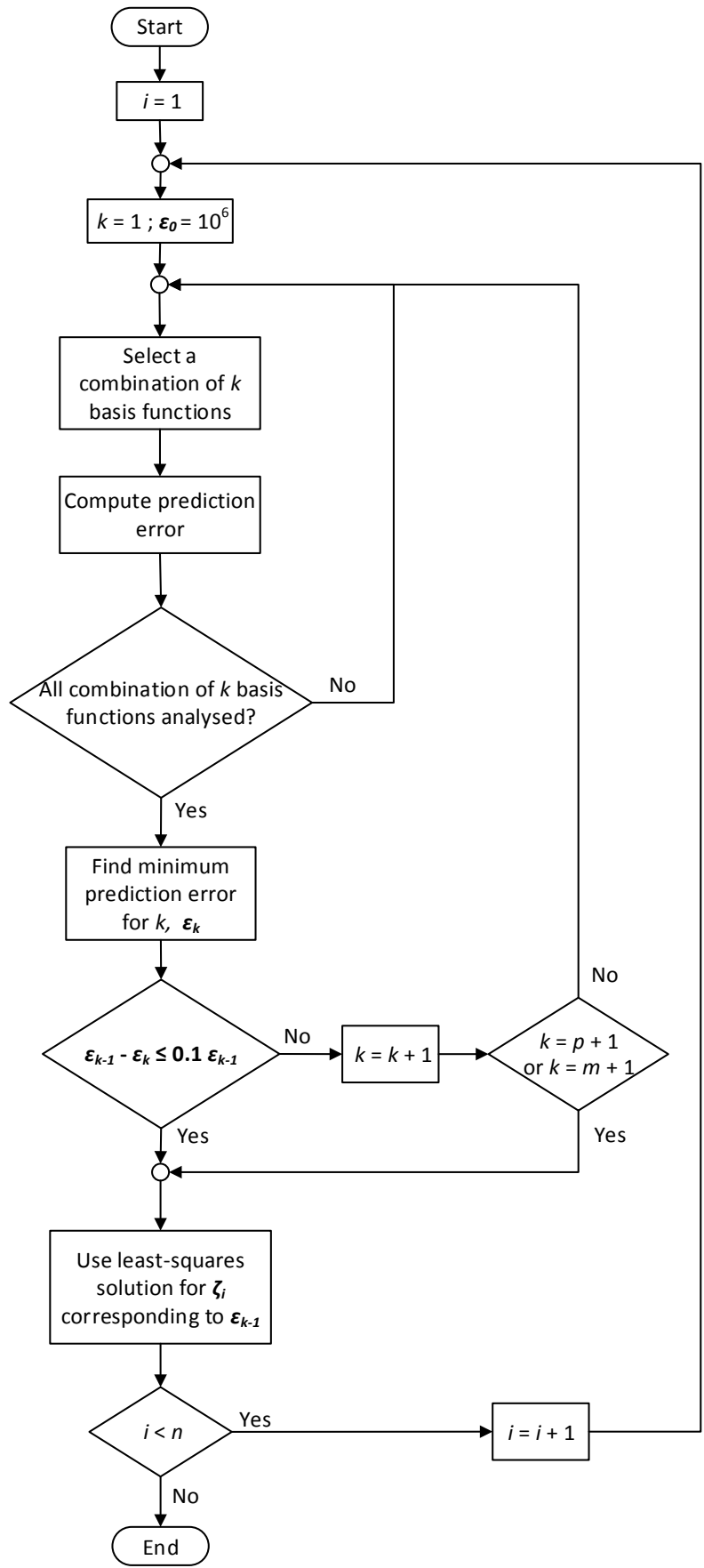

Figure 1 – Block diagram of the proposed approach to system identification based on subset growth (SISuG). The number of state variables is denoted by $n$, the number of basis functions by $p$ and the number of samples by $m$. The number of basis functions with a non-zero weight is denoted by $k$.

## 2.2. LOOCV Error Trick

Consider the typical least-squares problem

$$y \approx \Theta \zeta$$

where the solution for $\zeta$ is estimated by

$$\hat{\zeta} = (\Theta^T \Theta)^{-1} \Theta^T y = \Theta^\dagger y$$

and $\Theta^\dagger$ denotes the Moore-Penrose pseudoinverse of matrix $\Theta$.

To obtain a prediction error for such solution, we can use a leave-one-sample-out cross-validation (LOOCV) scheme and compute the mean squared error, $\varepsilon$. In fact, $\varepsilon$ can be directly computed as follows without the need for running LOOCV $m$ times ($m$ denotes the number of samples) [10], [6],

$$\varepsilon = \frac{1}{m} \sum_{j=1}^{m} \left( \frac{y(j) - \hat{y}(j)}{1 - \mathbf{H}(j,j)} \right)^2$$

where the hat-matrix $\mathbf{H}$ is given by

$$\mathbf{H} = \Theta \Theta^\dagger$$

and

$$\hat{y} = \Theta \hat{\zeta} = \Theta \Theta^\dagger y = \mathbf{H} y$$

## 3. Results and Discussion

In this section we illustrate the proposed approach (SISuG) on two examples: a 6-node feedback ring and the Van der Pol oscillator.

## 3.1. Example 1

The 6-node feedback ring dynamics are given by

$$\begin{bmatrix} \dot{x}_1 \\ \dot{x}_2 \\ \dot{x}_3 \\ \dot{x}_4 \\ \dot{x}_5 \\ \dot{x}_6 \end{bmatrix} = \begin{bmatrix} -1 & 0 & 0 & 0 & 0 & -1 \\ 1 & -1 & 0 & 0 & 0 & 0 \\ 0 & -1 & -1 & 0 & 0 & 0 \\ 0 & 0 & -1 & -1 & 0 & 0 \\ 0 & 0 & 0 & 1 & -1 & 0 \\ 0 & 0 & 0 & 0 & 1 & -1 \end{bmatrix} \begin{bmatrix} x_1 \\ x_2 \\ x_3 \\ x_4 \\ x_5 \\ x_6 \end{bmatrix}, \quad x(t_0) = \begin{bmatrix} 1 \\ 0 \\ 0 \\ 0 \\ 0 \\ 0 \end{bmatrix}$$

where $\mathbf{Z} \in \mathbb{R}^{6 \times 6}$, $x(t_0)$ is the initial condition and the following set of basis functions is considered, $\boldsymbol{\theta}^T = [x_1 \ x_2 \ x_3 \ x_4 \ x_5 \ x_6]$.

Let us consider the scenario where $m - 1$ samples are unevenly measured over 6 units of time at time-stamps randomly given as follows,

$$t_d = t_0 + d \cdot T + \Delta_t, \quad d = 1, \dots, m - 1$$

where

$$T = \frac{6}{m-1}; \quad \Delta_t \sim \mathcal{U}\left(-\frac{1}{4}T, \frac{1}{4}T\right)$$

and $\mathcal{U}$ denotes the continuous uniform distribution.

Different values of $m$ were tested and the root mean squared error ($RMSE$) was used to measure the quality of the obtained solution $\hat{\mathbf{Z}}$,

$$RMSE = \sqrt{\frac{1}{np} \sum_{i=1}^{n} \sum_{j=1}^{p} [\mathbf{Z}(i,j) - \hat{\mathbf{Z}}(i,j)]^2}$$

The $RMSE$ averaged over 200 simulations is plotted in Figure 2 (a) for different values of $m$. The results show that the method performs well in the case of unevenly sampled data and, as expected, the performance improves as the number of samples increases.

Analysing now the case where $m - 1 = 12$ and considering that the sampling period is constant over time ($\Delta_t = 0$), Figure 3 illustrates how the algorithm evolves as the number of basis functions, $k$, is increased. The obtained solution, $\widehat{\mathbf{Z}}$, had a $RMSE = 0.01$. Note that $k$ is correctly set at 2 basis functions for all state variables.

$$\widehat{\mathbf{Z}} = \begin{bmatrix} -0.993 & 0 & 0 & 0 & 0 & -0.995 \\ 0.962 & -0.960 & 0 & 0 & 0 & 0 \\ 0 & -0.985 & -0.986 & 0 & 0 & 0 \\ 0 & 0 & -1.001 & -1.001 & 0 & 0 \\ 0 & 0 & 0 & 0.999 & -0.999 & 0 \\ 0 & 0 & 0 & 0 & 1.000 & -1.000 \end{bmatrix}$$

It is also important to note that zeros in $\widehat{\mathbf{Z}}$ correspond to exact zeros, i.e., no threshold operation was used to round them to zero.

## 3.2. Example 2

We consider the dynamics of the Van der Pol oscillator,

$$\dot{x}_1 = x_2$$

$$\dot{x}_2 = -x_1 + x_2 - x_1^2 x_2$$

with initial conditions

$$x_1(t_0) = -1; \quad x_2(t_0) = 1$$

and the following set of basis functions,

$$\boldsymbol{\theta}^T = [x_1 \quad x_2 \quad x_1 x_2 \quad x_1^2 \quad x_2^2 \quad x_1^2 x_2 \quad x_2^2 x_1 \quad x_1^3 \quad x_2^3]$$

The same scenario where data was unevenly and randomly samples was considered. Different values of $m$ were tested and, the $RMSE$ averaged over 200 simulations is plotted in Figure 2 (b). Again, the results show that the method performs well in the case of unevenly sampled data and, as expected, the performance improves as the number of samples increases.

Analysing now the case where $m - 1 = 12$ and considering that the sampling period is constant over time ($\Delta_t = 0$), Figure 4 illustrates how the algorithm evolves as the number of basis functions, $k$, is increased. The obtained solution, $\widehat{\mathbf{Z}}$, had a $RMSE = 0.05$.

$$\widehat{\mathbf{Z}} = \begin{bmatrix} 0 & 0.975 & 0 & 0 & 0 & 0 & 0 & 0 & 0 \\ -0.971 & 0.830 & 0 & 0 & 0 & -0.864 & 0 & 0 & 0 \end{bmatrix}$$

Note that, despite the small number of samples, $k$ is correctly set at 1 basis function and 3 basis functions for $x_1$ and $x_2$, respectively.

## 4. Conclusion

In this work, we propose an approach to sparse continuous-time system identification when data is unevenly sampled, i.e. the sampling frequency changes randomly over time. The method was illustrated on two examples as a proof of concept. In future work, a deeper analysis should be conducted, considering more examples as well as directly comparing the results with other methods in the field of continuous-time system identification.

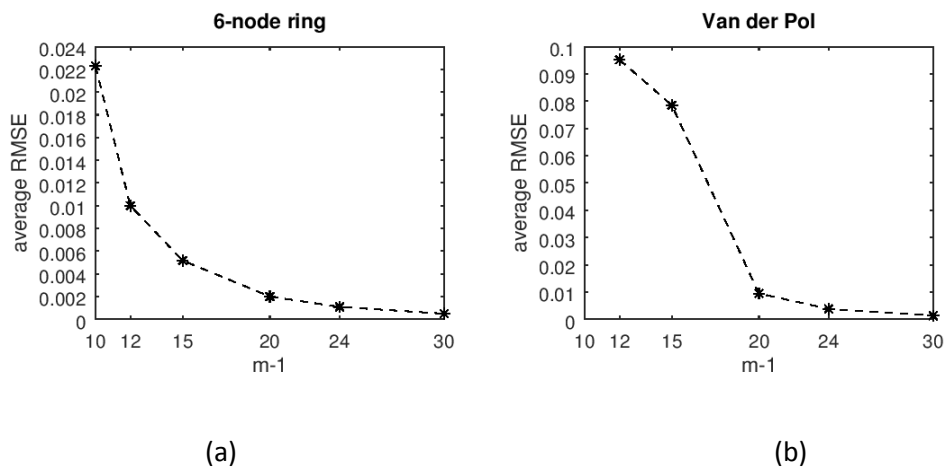

(a)  (b)

Figure 2 – Average $RMSE$ when $m - 1$ samples are unevenly and randomly measured over 6 units of time: (a) 6-node feedback ring, (b) Van der Pol oscillator. For each value of $m$, the $RMSE$ was averaged over 200 simulations.

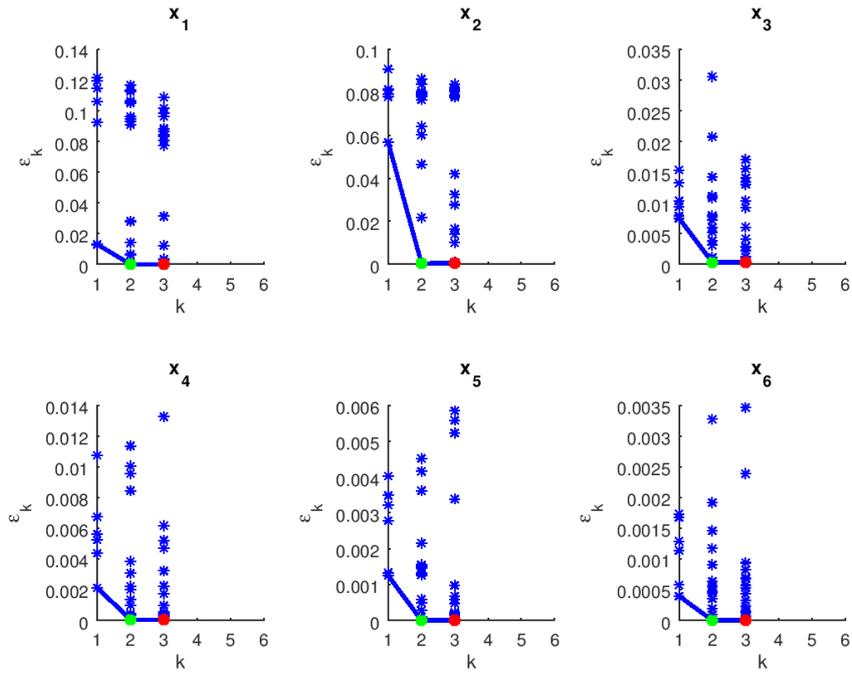

Figure 3 – 6-node feedback ring example ($m - 1 = 12$, $\Delta_t = 0$ ). It illustrates how SISuG algorithm evolves as the number of basis functions, $k$, is increased. Each point represents a different combination of the basis functions for each state variable $x_i$. The moment the iterative process stops is represented by the red dot. The combination of basis functions represented by the green dot is then selected.

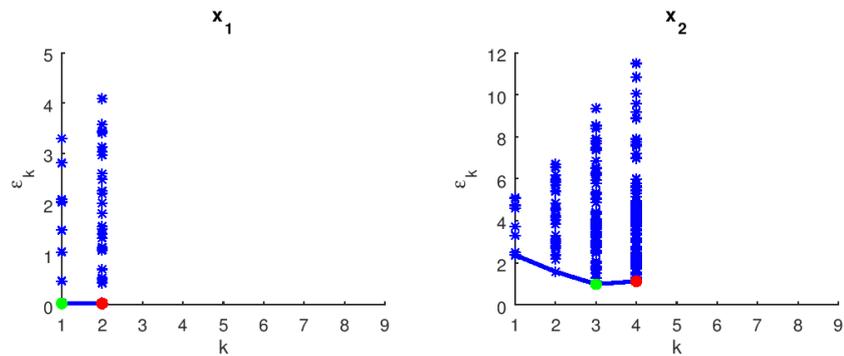

Figure 4 – Van der Pol oscillator example ($m - 1 = 12$, $\Delta_t = 0$ ). It illustrates how SISuG algorithm evolves as the number of basis functions, $k$, is increased. Each point represents a different combination of the basis functions for each state variable $x_i$. The moment the iterative process stops is represented by the red dot. The combination of basis functions represented by the green dot is then selected.